\documentclass[a4paper,11pt]{article}

\usepackage{jheppub} % for details on the use of the package, please
\usepackage[T1]{fontenc} % if needed

\usepackage{graphicx}
\usepackage{amssymb}
\usepackage{url}
\usepackage{subfigure}
\usepackage{array}
\usepackage{xspace}
\usepackage{enumerate}
\usepackage{atlasphysics}

\newcommand{\papertitle}{Search for direct chargino production in anomaly-mediated supersymmetry breaking models based on a disappearing-track signature in $pp$ collisions at $\sqrt{s}=7~\TeV$ with the ATLAS detector}

\def\chinoone{\ensuremath{\mathchoice%
      {\displaystyle\raise.4ex\hbox{$\displaystyle\tilde\chi_1$}}%
         {\textstyle\raise.4ex\hbox{$\textstyle\tilde\chi_1$}}%
       {\scriptstyle\raise.3ex\hbox{$\scriptstyle\tilde\chi_1$}}%
 {\scriptscriptstyle\raise.3ex\hbox{$\scriptscriptstyle\tilde\chi_1$}}}}

\def\mth{m_{3/2}}
\def\mdiff{\Delta m_{\chinoone}}
\def\nTRT{N_{\mathrm{TRT}}^{\mathrm{outer}}}
\def\ptem{\ensuremath{p_{\mathrm{T}}^{\mathrm{jet,EM}}}}

\def\trigdphi{\Delta \phi_{\mathrm{min}}^{\mathrm{jet-}\met}}

\def\nBlayer{N_{\mathrm{b-layer}}}
\def\nPixel{N_{\mathrm{pixel}}}
\def\nSCT{N_{\mathrm{SCT}}}
\def\probedis{\mathcal{P}^{\mathrm{dis}}_{e}}

\usepackage{preprintcover}  % load package
\PreprintCoverPaperTitle{\papertitle}  % paper title
\PreprintIdNumber{CERN-PH-EP-2012-243}  % CERN preprint number
\PreprintCoverAbstract{
A search for direct chargino production in anomaly-mediated supersymmetry breaking scenarios is performed in $pp$ collisions at $\sqrt{s} = 7 \TeV$ using 4.7 $\ifb$ of data collected with the ATLAS experiment at the LHC.
In these models, the lightest chargino is predicted to have a lifetime long enough to be detected in the tracking detectors of collider experiments.
This analysis explores such models by searching for chargino decays that result in tracks with few associated hits in the outer region of the tracking system.
The transverse-momentum spectrum of candidate tracks is found to be consistent with the expectation from the Standard Model background processes and constraints on chargino properties are obtained.
}
\PreprintJournalName{JHEP}  % journal name
\title{\boldmath \papertitle}

\author{The ATLAS collaboration}
\affiliation{CERN, 1211 Geneva 23, Switzerland}
\emailAdd{atlas.publication@cern.ch}

\abstract{
A search for direct chargino production in anomaly-mediated supersymmetry breaking scenarios is performed in $pp$ collisions at $\sqrt{s} = 7 \TeV$ using 4.7 $\ifb$ of data collected with the ATLAS experiment at the LHC.
In these models, the lightest chargino is predicted to have a lifetime long enough to be detected in the tracking detectors of collider experiments.
This analysis explores such models by searching for chargino decays that result in tracks with few associated hits in the outer region of the tracking system.
The transverse-momentum spectrum of candidate tracks is found to be consistent with the expectation from the Standard Model background processes and constraints on chargino properties are obtained.
}

\begin{document} 
\maketitle
\flushbottom

\section{Introduction}\label{sec:introduction}
Anomaly-mediated supersymmetry breaking (AMSB) models~\cite{Giudice:1998xp,Randall:1998uk}, where soft supersymmetry (SUSY) breaking is caused by loop effects, provides a constrained mass spectrum of SUSY particles.
In particular, the ratios of the three gaugino masses are given approximately by $M_1:M_2:M_3$ $\approx$ $3:1:7$ , where $M_i$ $(i=1,2,3)$ are the bino, wino and gluino masses, respectively.
The lightest gaugino is the wino, and the lightest chargino ($\chinoonepm$) and neutralino ($\ninoone$ as the lightest supersymmetric particle) are the charged and neutral winos. The mass of $\chinoonepm$ ($m_{\chinoonepm}$) becomes slightly heavier than that of $\ninoone$ due to radiative corrections involving electroweak gauge bosons.
The typical mass splitting between the charged and neutral winos ($\mdiff$) is $160\text{--}170 \MeV$\footnote{Throughout this paper, natural units are used, such that $\hbar = c = 1$.}. %, which is constant over most of the AMSB parameter space.
This phenomenological feature of the nearly degenerate $\chinoonepm$ and $\ninoone$ has the important implication that the $\chinoonepm$ has a considerable lifetime and predominantly decays into $\ninoone$ plus a low-momentum ($\sim 100 \MeV$) $\pi^{\pm}$.
The mean lifetime of the $\chinoonepm$ ($\tau_{\chinoonepm}$) is expected to be typically a fraction of a nanosecond. Therefore, some charginos will have decay lengths exceeding a few tens of centimeters at the energies of the Large Hadron Collider (LHC) and their tracks may have no or few associated hits in the outer region of the tracking system, causing them to be classified as ``disappearing tracks''.

This paper describes a search for the production of long-lived AMSB charginos, via electroweak processes in $pp$ collisions at $\rts = 7 \TeV$:
\begin{equation*}
  pp \rightarrow \chinoonepm \ninoone j, \hspace{12pt}  pp \rightarrow \chinoonep \chinoonem j.
\end{equation*}
with their subsequent decays, where $j$ denotes an energetic jet from initial-state radiation used to trigger the signal event.
Since the $\chinoonepm$ could decay in the inner tracking volume and the $\ninoone$ escapes from the detector, the resulting signal topology is characterized by a high-$\pt$ (transverse momentum) jet, large missing transverse momentum (its magnitude is denoted by $\met$), and a high-$\pt$ disappearing track.
A previous search for a disappearing-track signature~\cite{ATLAS:2012ab} by the ATLAS collaboration was based on signal production via the strong interaction, resulting in final states with multiple high-$\pt$ jets and large $\met$.
Given the ratio $M_3/M_2$ $\approx$7, the masses of coloured particles are comparatively large and thus the cross-sections are small compared to those from electroweak production.  

\section{The ATLAS detector}
ATLAS is a multi-purpose detector~\cite{Aad:2008zzm}, covering nearly the entire solid angle\footnote{ATLAS uses a right-handed coordinate system with its origin at the nominal interaction point (IP) in the centre of the detector and the $z$-axis coinciding with the axis of the beam pipe. The $x$-axis points from the IP to the centre of the LHC ring, and the $y$-axis points upward.  Cylindrical coordinates ($r,\phi$) are used in the transverse plane, $\phi$ being the azimuthal angle around the beam pipe. Pseudorapidity is defined in terms of the polar angle $\theta$ as $\eta = -\ln \tan(\theta/2)$.} around the collision point with layers of tracking devices surrounded by a superconducting solenoid providing a 2 tesla axial magnetic field, a calorimeter system, and a muon spectrometer.
The inner detector (ID) provides track reconstruction in the region $|\eta| < 2.5$ and consists of pixel and silicon microstrip (SCT) detectors inside a straw tube transition radiation tracker (TRT).
Of particular importance to this analysis is the TRT detector.
The barrel TRT covers the region $|z|$ $<$ $780~\mathrm{mm}$ and is divided into inner, middle, and outer concentric rings of 32 modules each, comprising a stack in azimuthal angle.
They cover the radial ranges $563~\mathrm{mm}$ to $694~\mathrm{mm}$ (inner), $697~\mathrm{mm}$ to $860~\mathrm{mm}$ (middle), and $863~\mathrm{mm}$ to $1066~\mathrm{mm}$ (outer).
A module consists of a carbon-fiber laminate shell and an array of straw tubes. %and has a different structure for each ring.
The average numbers of pixel, SCT and TRT hits on a track going through the inner detector in the central region are about 3, 8 and 34, respectively.
The calorimeter system covers the range $|\eta| < 4.9$.
The electromagnetic calorimeter is a lead/liquid-argon (LAr) detector in the barrel ($|\eta| < 1.475$) and endcap ($1.375 < |\eta| < 3.2$) regions.
The hadronic calorimeters are composed of a steel and scintillator barrel ($|\eta| < 1.7$), a LAr/copper endcap ($1.5 < |\eta| <3.2$), and a LAr forward system ($3.1< |\eta| <4.9$) with copper and tungsten absorbers.
The muon spectrometer consists of three large superconducting toroids, trigger chambers, and precision tracking chambers which provide muon momentum measurements up to $|\eta|$ of 2.7.

\section{Data and simulated event samples}\label{sec:samples}
This search is based on $pp$ collision data at $\rts = 7 \TeV$ recorded by the ATLAS detector in 2011, corresponding to an integrated luminosity of $4.7~\ifb$ after the application of beam, detector, and data quality requirements.

The large cross-section of QCD di-jet events especially at small $\pt$ is suppressed at the trigger level by requiring at least one jet with $\pt > 55 \GeV$, $\met > 55 \GeV$, and $\trigdphi > 1~\mathrm{rad}$, where $\trigdphi$ indicates the smallest azimuthal separation between the missing transverse momentum and either of the two highest-$\pt$ jets with $\pt > 30 \GeV$.
The jet $\pt$ and $\met$ for the trigger are based on calorimeter information and measured at the electromagnetic scale.
For background events $\met$ is usually aligned with a high-$\pt$ jet ($\trigdphi \approx 0~\mathrm{rad}$) since it is due to jet mis-measurements while for the signal $\trigdphi \approx \pi~\mathrm{rad}$ as it arises from the outgoing neutralinos.

Simulated Monte Carlo (MC) events are used to assess the experimental sensitivity to given models.
The minimal AMSB model is characterized by four parameters: the gravitino mass ($m_{3/2}$), the universal scalar mass ($m_0$), the ratio of Higgs vacuum expectation values at the electroweak scale ($\tan\beta$), and the sign of the higgsino mass term ($\mu$).
{\sc Isasusy} from {\sc Isajet} v7.80~\cite{Paige:2003mg} is used to calculate the SUSY mass spectrum and the decay tables.
The MC signal samples are produced using Herwig++~\cite{Bahr:2008pv} with {\sc MRST2007 LO*}~\cite{Sherstnev:2007nd} parton distribution functions (PDFs).
All simulated samples used in this paper are produced using a detector simulation based on {\sc Geant4} \cite{Agostinelli:2002hh,Aad:2010wqa}, and include multiple $pp$ interactions per event (pile-up) to model that observed in data.
Signal cross-sections are calculated at next-to-leading order (NLO) in the strong coupling constant using {\sc Prospino2}~\cite{Beenakker:1996ch}, as shown in figure~\ref{figure:xsec_EW}.
The nominal cross-section and its uncertainty are taken from an envelope of cross-section predictions using different PDF sets and factorisation and renormalisation scales, as described in ref.~\cite{Kramer:2012bx}.
Simulated points with chargino masses ranging from $70\text{--}300 \GeV$ are studied, and in particular two reference points with $m_{\chinoonepm}$ $\sim 100 \GeV$ and $200 \GeV$ are illustrated in this paper.
A large value of $1 \TeV$ is used for $m_0$ in order to prevent the existence of a tachyonic slepton. However, the production cross-section is determined only by the wino mass ($\propto m_{3/2})$, and the results presented in this paper are largely independent of the other parameters.
The mean lifetime $\tau_{\chinoonepm}$ is set to $1~\mathrm{ns}$, the value for which this analysis has the highest sensitivity.
Samples with different mean lifetimes are obtained by applying event weights to the original sample, such that the distribution of the proper lifetime follows that for a given mean lifetime value.
The branching fraction for the decay $\chinoonepm \to \ninoone \pi^{\pm}$ is set to 100\%.

\begin{figure}[ht!!]
  \centering
  \includegraphics[width=0.75\textwidth]{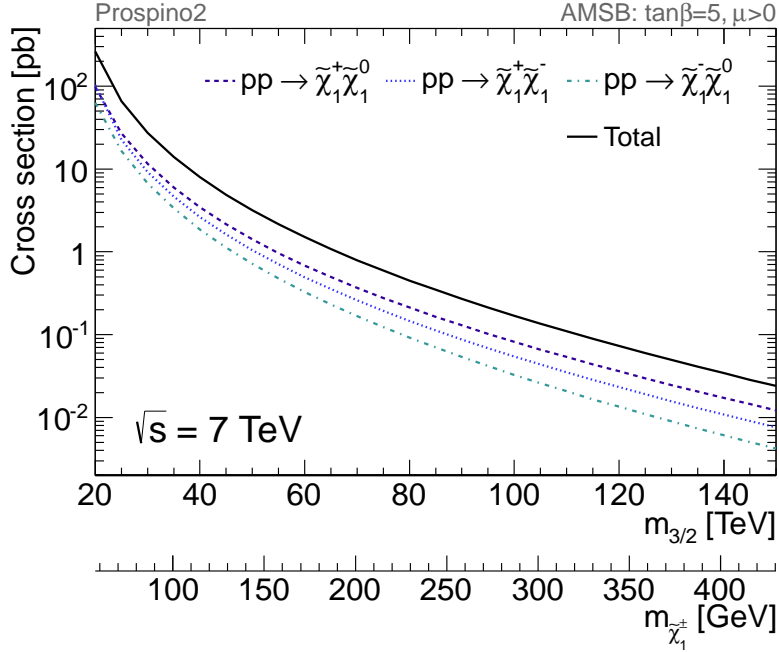}
  \caption{The cross-section for direct chargino production at $\rts = 7 \TeV$ as a function of $\mth$.
  The corresponding $m_{\chinoonepm}$ values for each $\mth$ are also indicated.}
  \label{figure:xsec_EW}
\end{figure}

\section{Event reconstruction and selection}\label{sec:selection}
Kinematic selection criteria are applied which ensure high trigger efficiency while reducing the Standard Model (SM) background arising from unidentified charged leptons that survive a lepton veto.
The vast majority of background events are removed by the TRT-based selection criteria that are used to identify the decay of the chargino.

\subsection{Event reconstruction} \label{sec:object}
The primary vertex~\cite{ATLAS-CONF-2010-069} is required to have at least five associated tracks; when more than one such vertex is found, the vertex with the largest total $|\pt|^2$ of the associated tracks is chosen.
Jets are reconstructed using the anti-$k_{t}$ algorithm~\cite{Cacciari:2008gp} with a distance parameter of 0.4. %in $y-\phi$ space.
The inputs to the jet reconstruction algorithm are topological calorimeter energy clusters.
The measurement of jet transverse momentum at the electromagnetic scale ($\ptem$) underestimates hadronic jets due to the nature of the non-compensating calorimeters and dead material.
Thus, an average correction depending on $\eta$ and $\ptem$. is applied to obtain the correct transverse momentum.
The details of the jet calibration procedure are given in ref.~\cite{Aad:2011he}.
In the analysis, requirements of $\pt > 20 \GeV$ and $|\eta| < 2.8$ are applied.
Electron candidates are selected with ``loose'' identification requirements, as described in ref.~\cite{Aad:2011mk} and required to fulfil the requirements of transverse energy, $\et > 10 \GeV$ and $|\eta| < 2.47$.
Muon candidates are identified by an algorithm which combines an ID track with either a track reconstructed in the muon spectrometer, or with a track segment in the innermost muon station~\cite{Aad:2008zzm,Aad:2011fp}.
Furthermore, muons are required to have at least one hit in the innermost layer of the pixel detector ($\nBlayer$) if crossing an active module of that layer, more than one pixel hit ($\nPixel$), at least six SCT hits ($\nSCT$), $\pt > 10 \GeV$ and $|\eta| < 2.4$.

Following the object reconstruction described above, overlaps between jets and leptons are resolved.
First, any jet candidate lying within a distance of $\Delta R \equiv \sqrt{\left( \Delta\eta \right)^2 + \left( \Delta\phi \right)^2} = 0.2$ of an electron is discarded.
Then, any lepton candidate within a distance of $\Delta R = 0.4$ of any surviving jet is discarded.

The calculation of $\met$ is based on the transverse momenta of jets and lepton candidates described above and all calorimeter energy clusters that are not associated to such objects~\cite{Aad:2012re}.

\subsection{Kinematic selection criteria} \label{sec:kinematic_selection}
Following the trigger decision, selection requirements to suppress non-collision background events, given in ref.~\cite{Aad:2011he}, are applied to jets.
Candidate events are then required to have no electron or muon candidates (lepton veto) to suppress the background events from $W/Z+\mathrm{jets}$ and top-quark pair-production processes.
The candidates are required to have $\met > 90 \GeV$ and at least one jet with $\pt > 90 \GeV$.
In order to further suppress the QCD background, $\trigdphi > 1.5~\mathrm{rad}$ for the two highest $\pt$ jets with $\pt > 50 \GeV$ is imposed.
The trigger selection is 98\% efficient for signal events satisfying these selection requirements.

\subsection{Disappearing-track selection criteria} \label{sec:track_selection}
The TRT detector provides substantial discrimination between penetrating and decaying charged particles: the average number of hits on a track going through the TRT in the barrel region is about 34 and consecutive hits can be observed along the track with small radial spacing between adjacent hits, while a smaller number is expected for charged particles that decay in the TRT volume.
If a chargino decays in the TRT volume, the track is still found with a high efficiency based on hits in the pixel and SCT detectors. 
Such a chargino track candidate can therefore be fully reconstructed by the ATLAS standard track reconstruction algorithm.

The tracks originating from charginos are expected to have high-$\pt$ and to be isolated.
Therefore, chargino candidate tracks are required to fulfil the following criteria:
\begin{enumerate}[{(}I{)}]
  \item The track must have $\nPixel \ge 1$, $\nBlayer \ge 1$ if crossing an active module of the innermost pixel layer, $\nSCT \ge 6$, $|d_0| < 1.5~\mathrm{mm}$ and $|z_0\sin\theta| < 1.5~\mathrm{mm}$, where $d_0$ and $z_0$ are the transverse and longitudinal impact parameters with respect to the primary vertex.
  \item The track must be isolated: there must be no tracks having $\pt$ above $0.4 \GeV$ within a cone of $\Delta R = 0.1$ around the candidate track. There must also be no jets having $\pt$ above $50 \GeV$ within a cone of $\Delta R = 0.4$.
  \item The candidate track must have $\pt$ above $10 \GeV$, and must be the highest-$\pt$ isolated track in the event.
  \item The relative uncertainty on the momentum measurement must be below 20\%.
  \item The candidate track must point to the TRT barrel layers but not the region around $|\eta| = 0$ ($0.1< |\eta| < 0.63$).
  \item The number of hits in the TRT outer module associated to the track ($\nTRT$) must be less than five.
\end{enumerate}
Criterion (I) is applied in order to ensure well-reconstructed primary tracks.
Criteria (II) and (III) are employed to select chargino tracks that are isolated and have the highest $\pt$ in most cases.
Tracks seeded from an incorrect combination of SCT space-points could have anomalously high values of $\pt$ and worse momentum resolution; criterion (IV) suppresses such tracks.
Criterion (V) is based on the extrapolated track position and is used to avoid inactive regions of the TRT and reject muons failing identification due to a small gap in acceptance around $\eta = 0$.
For criterion (VI), $\nTRT$ is calculated by counting TRT hits lying on the extrapolated track.
This criterion selects charginos decaying within the volume between the SCT outer layers and the TRT outer modules.
Hereafter, unless explicitly stated otherwise, ``high-$\pt$ isolated track selection'' and ``disappearing-track selection'' indicate criteria (I)--(V) and (I)--(VI), respectively.
Figure~\ref{figure:trt_outer} shows the $\nTRT$ distributions with the high-$\pt$ isolated track selection requirements for data, simulated signal MC events, and simulated MC SM background events.
Details of the SM background MC samples are described in ref.~\cite{ATLAS:0lep7tev5ifb}.
When charginos decay before reaching the TRT outer module, $\nTRT$ is expected to have a value near zero; conversely, charginos that reach the calorimeters and SM charged particles traversing the TRT typically have $\nTRT \simeq 15$.
The purity of chargino tracks in the signal MC events, defined as the fraction of candidate tracks matched to generated charginos, is almost 100\% at this stage; criterion (VI) removes the vast majority of background events.
Although it also reduces the signal efficiency, it strongly enhances the expected signal to background ratio.
A summary of kinematic selection criteria, disappearing-track requirements, and the data reduction are given in table~\ref{table:data_reduction}.
Signal efficiencies are low at the first stage due to the trigger based on initial-state radiation.
After the application of all selection criteria, 710 candidate events are selected.

\begin{figure}[ht!!]
  \centering
  \includegraphics[width=0.75\textwidth]{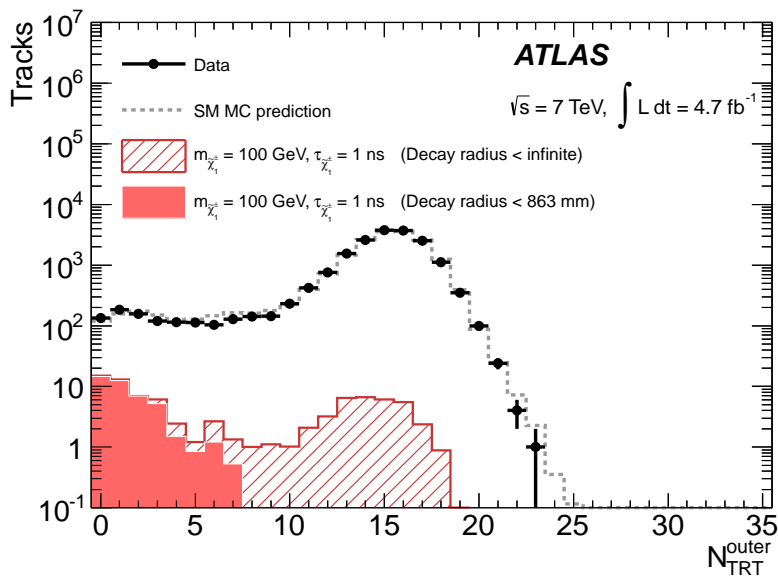}
  \caption{The $\nTRT$ distribution for data and signal events ($m_{\chinoonepm}$ = $100 \GeV$, $\tau_{\chinoonepm}$ = $1~\mathrm{ns}$) with the high-$\pt$ isolated track selection.
  The expectation from SM MC events, normalized to the number of observed events, is also shown.}
    \label{figure:trt_outer}
\end{figure}

\begin{table*}[ht!!]
  \centering
  \begin{tabular*}{\textwidth}{@{\extracolsep{\fill}}crrr@{}}
    \hline
    \hline
    Requirement & Observed & \multicolumn{2}{c}{Signal events (efficiency~[\%])} \\
    & & $m_{\chinoonepm} = 100 \GeV$ & $m_{\chinoonepm} = 200 \GeV$ \\
    \hline
    Quality requirements and trigger                           & 3765627   & 1983 (3.0)   & 283.3 (6.7) \\
    Non-collision background rejection                         & 2899498   & 1958 (3.0)   & 279.6 (6.6) \\
    Lepton veto                                                & 2186581   & 1906 (2.9)   & 274.8 (6.5) \\
    Leading jet $\pt > 90 \GeV$                            & 2054262   & 1497 (2.3)   & 237.7 (5.6) \\
    $\met > 90 \GeV$                                          & 1233864   & 1420 (2.2)   & 230.2 (5.5) \\
    $\trigdphi > 1.5~\mathrm{rad}$                                       & 1191298   & 1402 (2.1)   & 227.4 (5.4)  \\
    High-$\pt$ isolated track selection                        & 18493     & 90.5 (0.14)  & 9.1 (0.26) \\
    Disappearing-track selection                               & 710       & 42.9 (0.066) & 4.1 (0.12) \\
    \hline
    \hline
  \end{tabular*}
  \caption{Summary of selection requirements and data reduction for data and expected signal events ($\tau_{\chinoonepm}$ = $1~\mathrm{ns}$). The signal selection efficiencies are also shown in parentheses. }
  \label{table:data_reduction}
\end{table*}

\section{Estimate of the $\pt$ spectrum of the background contributions} \label{sec:background}
According to MC simulation, the background contribution after the high-$\pt$ isolated track selection comes predominantly from the $W(\rightarrow\tau\nu)$+jets production in which $\tau$ decay products fulfil the selection criteria.
Sub-leading contributions to the background come from prompt electrons failing to satisfy their identification criteria.
A background estimation based on the MC simulation suffers from large uncertainties due to low numbers of tracks after all the selection requirements and has difficulty in simulating the properties of these background tracks.
Therefore, an approach using data-driven control samples enriched in these background categories is employed to estimate the background track $\pt$ spectrum.
A simultaneous fit is then performed for signal and background yields using the $\pt$ spectrum of observed tracks.

\subsection{Interacting charged hadrons}
High-$\pt$ charged hadrons (mostly charged pions) can interact with material in the TRT detector and some tracks can be labelled as disappearing tracks; according to MC simulation, they are responsible for more than 80\% of the background in the signal search sample.
The $\pt$ spectrum of interacting hadron tracks is obtained from that of non-interacting hadron tracks, in a data-driven way using a data sample enriched in this background category as described in ref.~\cite{ATLAS:2012ab}.
In the $\pt$ range above $10 \GeV$, where inelastic interactions dominate, the interaction rate has nearly no $\pt$-dependence~\cite{Nakamura:2010zzi}.
By adopting the same kinematic selection criteria as those for the signal and ensuring a penetration through the TRT detector by requiring $\nTRT>10$, a sample of high-$\pt$ non-interacting hadron tracks is obtained.
The contamination from electron tracks and any chargino signal is removed by requiring the associated calorimeter activity, $\ET^{\mathrm{cone20}}/\pt^{\mathrm{track}}$, to be larger than 0.2, where $\ET^{\mathrm{cone20}}$ is the calorimeter transverse energy deposited in a cone of $\Delta R =  0.2$ around the track, excluding $\ET$ of its corresponding calorimeter cluster, and $\pt^{\mathrm{track}}$ is the track $\pt$.
According to MC simulation, the purity of non-interacting hadron tracks is $>99\%$ after these requirements.
These hadron tracks have a steeply falling $\pt$ spectrum, as shown in figure~\ref{figure:hadtrack}.
An ansatz functional form ${(1+x)^{a_0}}/{x^{a_1+a_2\ln(x)}}$ is then fitted to the $\pt$ spectrum of the control sample, where $x\equiv \pt^{\mathrm{track}}$ and $a_{i}$ $(i=0,1,2)$ are the fitted parameters.
The data distribution is well described by this functional form; a $\chi^2$ per degree of freedom (DOF) of 39/50 is calculated from the difference between the data and the best-fit form.

\begin{figure}[ht!!]
  \centering
  \includegraphics[width=0.70\textwidth]{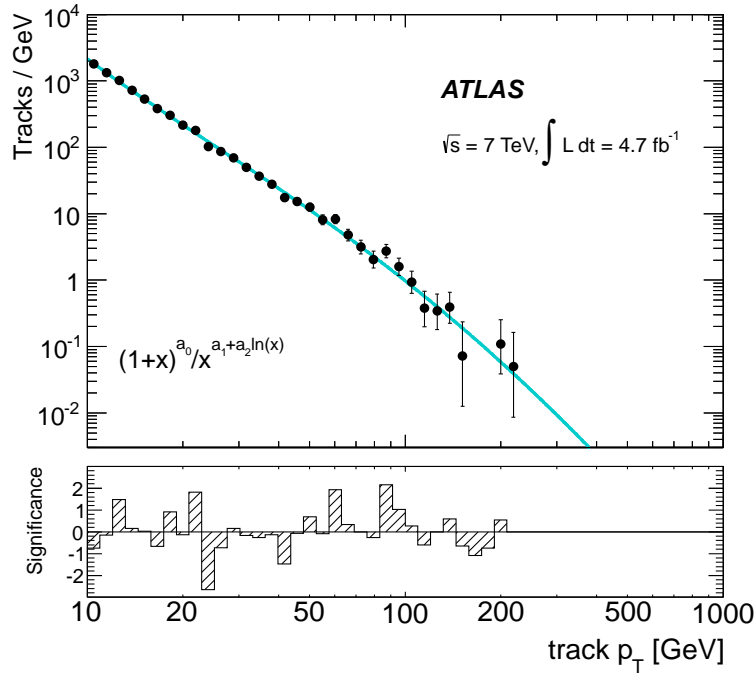}
  \caption{The $\pt$ distribution of the hadron-track background control sample.  The data and the fitted shape are shown by solid circles and a line, respectively.
  The significance of the residuals between the data and the fit model on a bin-by-bin basis is shown at the bottom of the figure.}
  \label{figure:hadtrack}
\end{figure}

\subsection{Electrons failing to satisfy identification criteria} \label{sec:electron}
The charged lepton background is mostly due to large bremsstrahlung where, predominantly, low-$\pt$ electrons contribute to this background.
Muons failing to satisfy the identification criteria could be also classified as disappearing tracks; however, this contribution is negligibly small since the probability of bremsstrahlung photon emission is proportional to $1/m_{\ell}^2$, where $m_{\ell}$ is the lepton mass.

In order to estimate the electron background, a control sample is defined by requiring the same kinematic selection requirements as for the signal search sample, but requiring one electron that fulfils ``medium'' identification criteria~\cite{Aad:2011mk} and the isolated track selection criteria; the purity of electrons is close to 100\% according to MC simulation.
The $\pt$ spectrum of electrons without any identification requirements is obtained by applying the correction for the medium identification efficiency~\cite{Aad:2011mk}.
This efficiency depends on $\pt$ and $\eta$, with an average value around 0.8.
The $\pt$ distribution of electron background tracks is then estimated by multiplying the corrected distribution (described above) by the probability of failing to satisfy the loose identification criteria (hence being retained in the signal search sample) and passing the disappearing-track selection criteria for electrons ($\probedis$).
For the measurement of $\probedis$, a ``tag-and-probe'' method is applied to $Z \to ee$ events collected with unprescaled single-electron triggers.
In order to ensure a very pure sample of $Z \to ee$ events, tag-electrons must be well isolated from jets and also required to fulfil ``tight'' identification criteria~\cite{Aad:2011mk} and have $\et > 25 \GeV$.
First, the $Z \to ee$ sample is selected by requiring no identified muons, at least one tag-electron and one high-$\pt$ isolated track.
Probe-electrons are selected without any identification requirements but with exactly the same high-$\pt$ isolated track selection criteria used for chargino candidate tracks.
Then, the reconstructed invariant mass is required to be within the range from $85\text{--}95 \GeV$; its value is calculated using the calorimeter energy for the tag and the track momentum for the probe.
The track momentum is used for the probe electron, since in the absence of any electron identification the precise calorimeter energy is not well defined.
The probability $\probedis$ is finally given by the fraction of events in which the probe-electron passes the disappearing-track selection criteria; it ranges from $10^{-2}$ to $10^{-4}$ for $10 < \pt < 50 \GeV$.
Due to too few data events, the nominal values of $\probedis$ are derived using MC events; no visible dependence on $\pt$ is found, and the average $\probedis$ values for data and MC events agree within 13\%, which is taken as a systematic uncertainty.

Figure~\ref{figure:eletrack} shows the resulting $\pt$ spectrum of electron background tracks; the systematic uncertainties on the identification efficiency are included.
The $\pt$-dependent identification efficiency and $\probedis$ produce a complicated spectrum; therefore, the electron background shape is determined by a fit to an extended functional form ${(x+b_0)^{b_1}}/{(x+b_2)^{b_3+b_4\ln(x)}}$ where $x\equiv \pt^{\mathrm{track}}$ and $b_{i}$ $(i=0,1,2,3,4)$ are the fitted parameters.
The $\chi^2$ per DOF is calculated to be 45/29.
Using this function the number of electron background tracks in the signal search sample is estimated to be $115 \pm 15$.
Statistical errors and uncertainties on the identification efficiency and $\probedis$ are considered in deriving the results.

\begin{figure}[ht!!]
  \centering
  \includegraphics[width=0.70\textwidth]{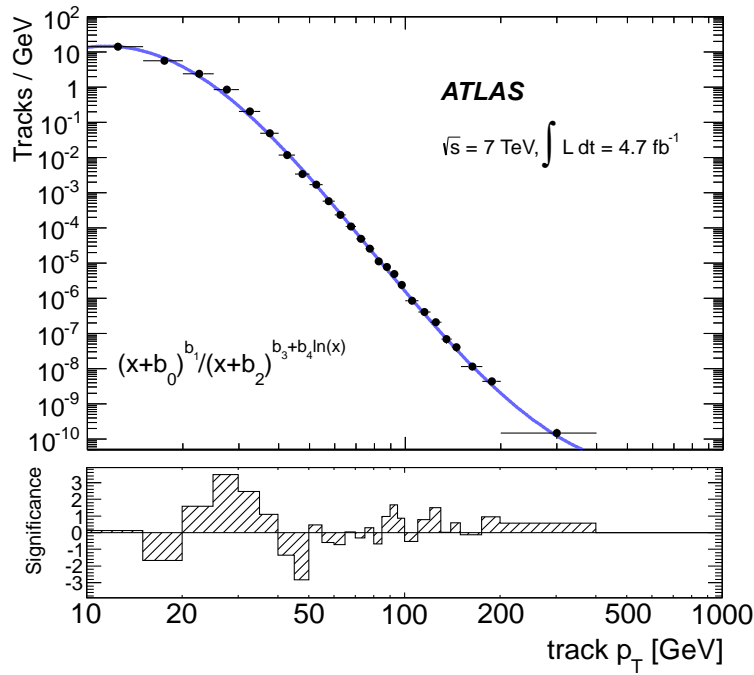}
  \caption{The estimated $\pt$ distribution of electron background tracks. The data and the fitted shape are shown by solid circles and a line, respectively.
  The error bars representing statistical errors and uncertainties on the identification efficiency are invisibly small.
  The significance of the residuals between the data and the fit model on a bin-by-bin basis is shown at the bottom of the figure.}
  \label{figure:eletrack}
\end{figure}

\section{Estimate of systematic uncertainties}\label{sec:uncertainties}
The sources of systematic uncertainty on the signal expectation which have been considered are the:
theoretical cross-section, parton radiation model, jet energy scale (JES) and resolution (JER), trigger efficiency, pile-up modelling, track reconstruction efficiency, and the integrated luminosity.

Theoretical uncertainties on the signal cross-section, already described in section~\ref{sec:samples}, range from 6--8\% depending on $m_{\chinoonepm}$.
High-$\pt$ jets originating from initial- and final-state radiation alter the signal acceptance.
The uncertainties on these processes are estimated by varying generator tunes in the simulation as well as by generator-level studies with an additional jet in the matrix-element method using {\sc MadGraph5}~\cite{Alwall:2011uj}+{\sc Pythia6}~\cite{Sjostrand:2006za}, after applying the kinematic selection criteria.
By adopting PDF tunes that provide less and more radiation and taking the maximum deviation from the nominal one, the uncertainty due to jet radiation is evaluated.
The uncertainty arising from the matching of matrix elements with parton showers is found by doubling and halving the default value of the matching parameter~\cite{Alwall:2007fs}.
The resulting changes are combined in quadrature and yield an uncertainty of 10--15\% depending on $m_{\chinoonepm}$.
The uncertainties on the JES and JER result in a variation of the signal selection efficiency; the variation of the signal selection efficiency arising from these uncertainties is assessed according to ref.~\cite{Aad:2011he}, and an uncertainty of 5--10\% is assigned.
An uncertainty of $3\%$ on the trigger efficiency is assigned by taking the difference between data and MC $W \to \mu\nu$ samples.
The uncertainty originating from the pile-up modelling in the simulation is evaluated by weighting simulated samples so that the average number of pile-up interactions is increased or decreased by 10\%; an uncertainty of $0.5\%$ is assigned.
The ID material affects the track reconstruction efficiency and the uncertainty due to the material description in the MC simulation is assessed as described in ref.~\cite{Aad:2011qe}.
By comparing the track reconstruction efficiency to that obtained with the MC samples with an extra 10\% of material in the tracking system, an uncertainty of $2\%$, in particular for tracks in the region of $|\eta| < 0.63$, is assigned.
The absolute luminosity of $pp$ collisions is determined with an uncertainty of $3.9\%$ \cite{Aad:2011dr,LuminosityGroup:1367408}.
The contributions of each systematic uncertainty in the signal expectation are summarized in table~\ref{table:summary_uncertainty} for the two reference signal samples.

Systematic uncertainties on the background are determined from the statistical uncertainties on the fit parameters and the full correlation matrix.
In addition, the $13\%$ uncertainty on the disappearing-track probability for electrons is considered (see section~\ref{sec:electron}).
Alternative fit functions for the $\pt$ shapes of the electron and interacting hadron tracks are also checked, showing that these agree with each other and with the original form within the fit uncertainties.
The effect on the sensitivity to the signals due to the choice of functional forms is thus found to be negligible.

\begin{table}[ht!!]
  \centering
  \begin{tabular*}{\linewidth}{@{\extracolsep{\fill}}lcc@{}}
    \hline
    \hline
    Source & $m_{\chinoonepm} = 100 \GeV$~[\%] & $m_{\chinoonepm} = 200 \GeV$~[\%] \\
    \hline
    \multicolumn{3}{l}{(\bf{Theoretical uncertainty})} \\
    Cross section       & 7 & 7 \\
    \hline
    \multicolumn{3}{l}{(\bf{Uncertainty on the acceptance})} \\
    Modeling of initial/final-state radiation & 10 & 13 \\
    JES/JER                                   & 10 & 6 \\
    Trigger efficiency                        & 3 & 3 \\
    Pile-up modelling                         & 0.5 & 0.5 \\
    Track reconstruction efficiency           & 2 & 2 \\
    Luminosity                                & 3.9 & 3.9 \\
    {\bf Sub-total} & 15 & 15 \\
    \hline
    \hline
  \end{tabular*}
  \caption{Summary of systematic uncertainties [\%] on the expectation of signal events.}
  \label{table:summary_uncertainty}
\end{table}

\section{Statistical analysis} \label{sec:statana}
In order to evaluate how well the observed data agree with a given signal model, a statistical test is performed based on maximizing a likelihood.
The likelihood function for the track $\pt$ in a sample of observed events ($n_{\mathrm{obs}}$) is defined as
\begin{equation}
  \prod^{n_{\mathrm{obs}}} \frac{ n_{\mathrm{s}}\mathcal{F}_{\mathrm{s}}(\pt) + n_{\mathrm{h}}\mathcal{F}_{\mathrm{h}}(\pt) + n_{\mathrm{e}}\mathcal{F}_{\mathrm{e}}(\pt)}{n_{\mathrm{s}} + n_{\mathrm{h}} + n_{\mathrm{e}}} \times \mathcal{L}_{\mathrm{sys}}, \\
\end{equation}
where $n_{\mathrm{s}}$, $n_{\mathrm{h}}$ and $n_{\mathrm{e}}$ are the number of signal events for a given value of the chargino mass and lifetime, the number of interacting hadron track events, and the number of electron track events, respectively.
The probability density function of the signal ($\mathcal{F}_{\mathrm{s}}$) is defined for a given value of the chargino mass and lifetime, and that of the interacting hadron (electron) tracks, $\mathcal{F}_{\mathrm{h}}$ ($\mathcal{F}_{\mathrm{e}}$), is shown in figure~\ref{figure:hadtrack} (\ref{figure:eletrack}).
In the fit, $n_{\mathrm{e}}$ is constrained to be its estimated value (see section~\ref{sec:electron}).
The effects of systematic uncertainties on the normalizations and the shape parameters describing the two $\pt$ distributions of the background tracks are incorporated via the constraining terms, $\mathcal{L}_{\mathrm{sys}}$, representing the product of normal and multivariate-normal distributions in which the variances are set to their uncertainties.

Figure~\ref{figure:trackpt_bestfit} shows the $\pt$ distribution for the selected data events compared to the background model obtained by the ``background-only'' fit in the $\pt$ range above $10 \GeV$.
The best-fit values of $n_{\mathrm{h}}$ and $n_{\mathrm{e}}$ are $610 \pm 30$ and $105 \pm 13$, respectively.
The probability of the fit to describe the data is 0.54. The numbers of expected background and observed tracks in the region $\pt > 50 \ (100) \GeV$ are $14.8 \pm 0.3$ and 19 ($2.20 \pm 0.05$ and 1), respectively, exhibiting no significant excess in the data.
The selected examples for the signal are also shown in figure~\ref{figure:trackpt_bestfit}.
The values of $n_{\mathrm{s}}$ for them, derived from the ``signal + background'' fit, are found to be consistent with zero.

\begin{figure}[ht!!]
  \centering
  \includegraphics[width=0.8\textwidth]{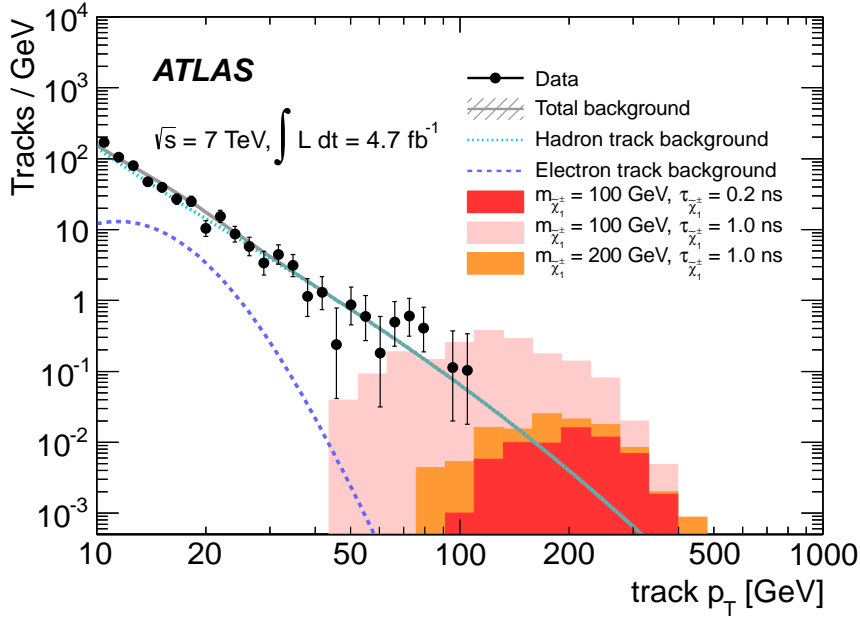}
  \caption{The $\pt$ distribution of candidate tracks. The solid circles show data and lines show background shapes obtained using the ``background-only'' fit. The contributions of two background components and the signal expectations are also shown. }
  \label{figure:trackpt_bestfit}
\end{figure}

\section{Results}\label{sec:results}
In the absence of a signal, constraints on $m_{\chinoonepm}$ and $\tau_{\chinoonepm}$ are set.
The upper limit on the production cross-section for a given $m_{\chinoonepm}$ and $\tau_{\chinoonepm}$ at 95\% confidence level (CL) is set by a point where the CL of the ``signal+background'' hypothesis, based on the profile likelihood ratio~\cite{Cowan:2010js} and the $CLs$ prescription~\cite{Read:2002hq}, falls below 5\% when scanning the CL along various values of signal strength.
The constraint on the $\tau_{\chinoonepm}\text{--}m_{\chinoonepm}$ parameter space is shown in figure~\ref{figure:limit_lifetime_vs_mass}.
The expected limit is set by the median of the distribution of 95\% CL limits calculated by pseudo-experiments with the expected background and no signal.
The expected number of background events is derived from the background-only fit in the region $10 < \pt < 50 \GeV$, where the systematic parameters are varied according to their systematic uncertainties when generating the ensemble of pseudo-experiments.

Figure~\ref{figure:limit_dM_vs_mass} shows the constraint on the $\mdiff\text{--}m_{\chinoonepm}$ parameter space of the minimal AMSB model.
The limits on $\tau_{\chinoonepm}$ are converted into limits on $\mdiff$ following ref.~\cite{Chen:1999yf}.
The region excluded by the LEP2 searches~\cite{Heister:2002mn,Abbiendi:2002vz,Abdallah:2003gv,LEP2} is also indicated.
For $\mdiff = 160 \ (170) \MeV$ ($\tau_{\chinoonepm} \sim 0.3~\mathrm{ns}$), the value most probable in the model, a new limit of $m_{\chinoonepm}> 103 \ (85) \GeV$ at 95\% CL is obtained.
For $\mdiff \sim 140 \MeV$, a more stringent limit of $m_{\chinoonepm} > 260 \GeV$ is set.

The analysis is not performed for signals having $\tau_{\chinoone} > 10~\mathrm{ns}$ (corresponding $\Delta m_{\chinoone}$ being below the charged pion mass) because a significant fraction of charginos would traverse the ID before decaying, thereby reducing the event selection efficiency.
These scenarios are considered as `stable'.

\begin{figure}[ht!!]
  \centering
  \includegraphics[width=0.75\textwidth]{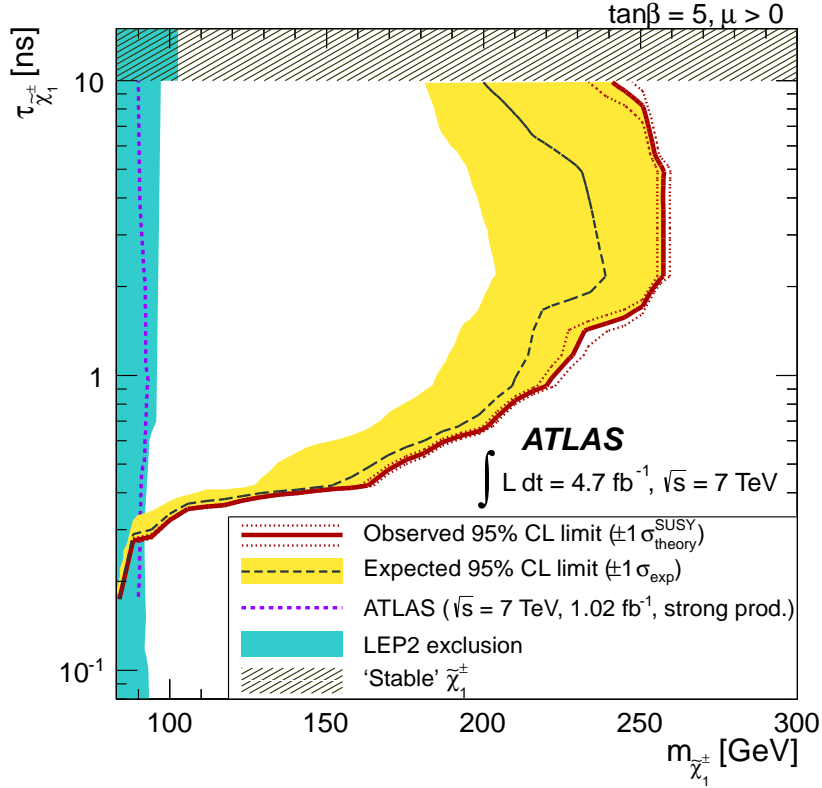}
  \caption{The constraint on the $\tau_{\chinoonepm}\text{--}m_{\chinoonepm}$ space for $\tan\beta = 5$ and $\mu>0$.
  The black dashed line shows the expected limits at 95\% CL, with the surrounding shaded bands indicating the 1$\sigma$ exclusions due to experimental uncertainties.
  Observed limits are indicated by the solid bold contour representing the nominal limit and the dotted lines on either side are obtained by varying the cross-section by the theoretical scale and PDF uncertainties.
  The previous result from ref.~\cite{ATLAS:2012ab} and the combined LEP2 exclusion at 95\% CL are also shown on the left by the dotted line and the shaded region, respectively.}
  \label{figure:limit_lifetime_vs_mass}
\end{figure}

\begin{figure}[ht!!]
  \centering
  \includegraphics[width=0.75\textwidth]{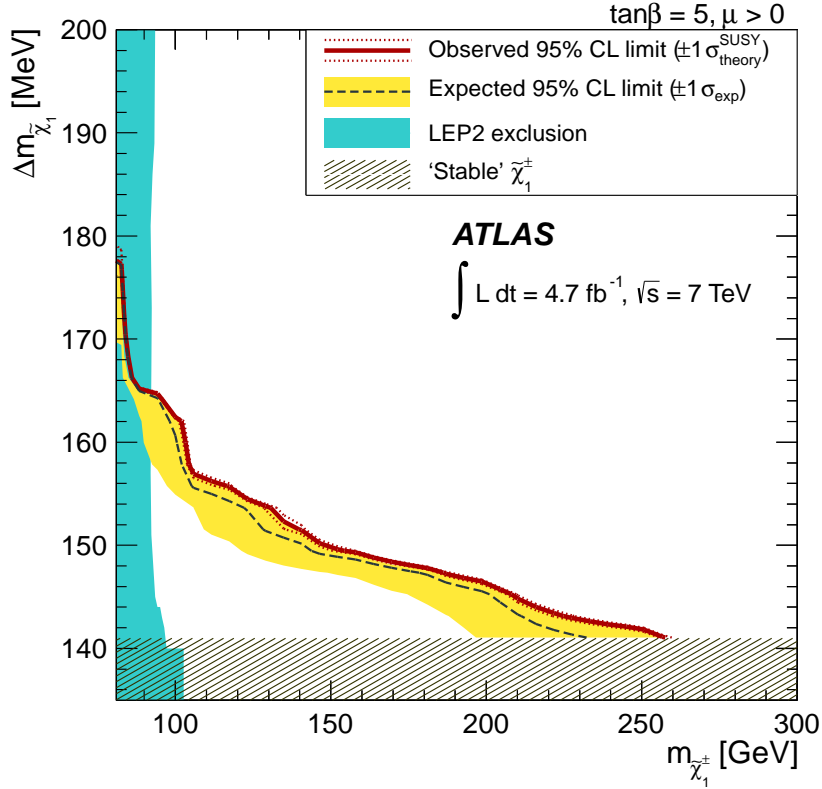}
  \caption{The constraint on the $\mdiff\text{--}m_{\chinoonepm}$ space of the AMSB model for $\tan\beta = 5$ and $\mu>0$, where $\tau_{\chinoonepm}$ is varying as described in figure~\ref{figure:limit_lifetime_vs_mass}.
  The dashed line shows the expected limits at 95\% CL, with the surrounding shaded bands indicating the 1$\sigma$ exclusions due to experimental uncertainties.
  Observed limits are indicated by the solid bold contour representing the nominal limit and the dotted lines on either side are obtained by varying the cross-section by the theoretical scale and PDF uncertainties.
  The combined LEP2 exclusion at 95\% CL is also shown on the left by the shaded region.
  Charginos in the lower shaded region could have significantly longer lifetime values for which this analysis has no sensitivity.}
  \label{figure:limit_dM_vs_mass}
\end{figure}

\section{Conclusions}\label{sec:conclusion}
The results of a search for the direct production of long-lived charginos in $pp$ collisions with the ATLAS detector using $4.7~\ifb$ of data have been presented in the context of AMSB scenarios.
The search is based on the signature of a high-$\pt$ isolated track with few associated hits in the outer part of the ATLAS tracking system, arising from a chargino decay into a neutralino and a low-$\pt$ pion.
The $\pt$ spectrum of observed candidate tracks is found to be consistent with the expectation from SM background processes.
Constraints on the chargino mass and the mass splitting between the lightest chargino and neutralino are set. A chargino having a mass below $103 \ (85) \GeV$ with a mass splitting of $160 \ (170) \MeV$, the most favoured scenario in the AMSB model, is excluded at 95\% CL.
This analysis provides a result complementary to the previous search based on signal production via the strong interaction~\cite{ATLAS:2012ab} and improves the sensitivity.
It also provides a largely AMSB-model-independent constraint on the chargino properties.
From the viewpoint of self-annihilating dark matter, a wino-like lightest SUSY particle with a mass of $\mathcal{O}(100) \GeV$ as obtained in certain AMSB scenarios, which simultaneously explains the observations by PAMELA~\cite{Adriani:2008zr} and Fermi LAT~\cite{FermiLAT:2011ab} as well as the WMAP relic density data~\cite{Jarosik:2010iu}, is of particular interest; it could be addressed with an increased LHC energy, more integrated luminosity and an extension of the analysis using shorter tracks.

\acknowledgments
We thank CERN for the very successful operation of the LHC, as well as the
support staff from our institutions without whom ATLAS could not be
operated efficiently.

We acknowledge the support of ANPCyT, Argentina; YerPhI, Armenia; ARC,
Australia; BMWF and FWF, Austria; ANAS, Azerbaijan; SSTC, Belarus; CNPq and FAPESP,
Brazil; NSERC, NRC and CFI, Canada; CERN; CONICYT, Chile; CAS, MOST and NSFC,
China; COLCIENCIAS, Colombia; MSMT CR, MPO CR and VSC CR, Czech Republic;
DNRF, DNSRC and Lundbeck Foundation, Denmark; EPLANET and ERC, European Union;
IN2P3-CNRS, CEA-DSM/IRFU, France; GNSF, Georgia; BMBF, DFG, HGF, MPG and AvH
Foundation, Germany; GSRT, Greece; ISF, MINERVA, GIF, DIP and Benoziyo Center,
Israel; INFN, Italy; MEXT and JSPS, Japan; CNRST, Morocco; FOM and NWO,
Netherlands; BRF and RCN, Norway; MNiSW, Poland; GRICES and FCT, Portugal; MERYS
(MECTS), Romania; MES of Russia and ROSATOM, Russian Federation; JINR; MSTD,
Serbia; MSSR, Slovakia; ARRS and MVZT, Slovenia; DST/NRF, South Africa;
MICINN, Spain; SRC and Wallenberg Foundation, Sweden; SER, SNSF and Cantons of
Bern and Geneva, Switzerland; NSC, Taiwan; TAEK, Turkey; STFC, the Royal
Society and Leverhulme Trust, United Kingdom; DOE and NSF, United States of
America.

The crucial computing support from all WLCG partners is acknowledged
gratefully, in particular from CERN and the ATLAS Tier-1 facilities at
TRIUMF (Canada), NDGF (Denmark, Norway, Sweden), CC-IN2P3 (France),
KIT/GridKA (Germany), INFN-CNAF (Italy), NL-T1 (Netherlands), PIC (Spain),
ASGC (Taiwan), RAL (UK) and BNL (USA) and in the Tier-2 facilities
worldwide.

\providecommand{\href}[2]{#2}\begingroup\raggedright

\clearpage 
% ATLAS Collaboration author list
% Data extracted on 30-Nov-2012 for paper reference SUSY-2012-19
%\documentclass[11pt]{article}
%\usepackage{a4wide}\begin{document}
\begin{flushleft}
{\Large The ATLAS Collaboration}

\bigskip

G.~Aad$^{\rm 48}$,
T.~Abajyan$^{\rm 21}$,
B.~Abbott$^{\rm 111}$,
J.~Abdallah$^{\rm 12}$,
S.~Abdel~Khalek$^{\rm 115}$,
A.A.~Abdelalim$^{\rm 49}$,
O.~Abdinov$^{\rm 11}$,
R.~Aben$^{\rm 105}$,
B.~Abi$^{\rm 112}$,
M.~Abolins$^{\rm 88}$
% ...
\bigskip
\\
% ...
$^{11}$ Institute of Physics, Azerbaijan Academy of Sciences, Baku, Azerbaijan\\
$^{12}$ Institut de F{\'\i}sica d'Altes Energies and Departament de F{\'\i}sica de la Universitat Aut{\`o}noma de Barcelona and ICREA, Barcelona, Spain\\
% ...
$^{21}$ Physikalisches Institut, University of Bonn, Bonn, Germany\\
% ...
$^{48}$ Fakult{\"a}t f{\"u}r Mathematik und Physik, Albert-Ludwigs-Universit{\"a}t, Freiburg, Germany\\
$^{49}$ Section de Physique, Universit{\'e} de Gen{\`e}ve, Geneva, Switzerland\\
% ...
$^{88}$ Department of Physics and Astronomy, Michigan State University, East Lansing MI, United States of America\\
% ...
$^{105}$ Nikhef National Institute for Subatomic Physics and University of Amsterdam, Amsterdam, Netherlands\\
% ...
$^{111}$ Homer L. Dodge Department of Physics and Astronomy, University of Oklahoma, Norman OK, United States of America\\
$^{112}$ Department of Physics, Oklahoma State University, Stillwater OK, United States of America\\
% ...
$^{115}$ LAL, Universit{\'e} Paris-Sud and CNRS/IN2P3, Orsay, France\\
% ...
\end{flushleft}

%\end{document}
% Created with ./xml2latex.py 


\begin{thebibliography}{10}

\bibitem{Giudice:1998xp}
G.~F. Giudice, M.~A. Luty, H.~Murayama, and R.~Rattazzi, {\it {Gaugino Mass  without Singlets}},  {\em JHEP} {\bf 12} (1998) 027,  [\href{http://xxx.lanl.gov/abs/hep-ph/9810442}{{\tt hep-ph/9810442}}].

\bibitem{Randall:1998uk}
L.~Randall and R.~Sundrum, {\it {Out of this world supersymmetry breaking}},  {\em Nucl. Phys.} {\bf B557} (1999) 79--118,  [\href{http://xxx.lanl.gov/abs/hep-th/9810155}{{\tt hep-th/9810155}}].

\bibitem{ATLAS:2012ab}
{\bf ATLAS} Collaboration, {\it {Search for anomaly-mediated supersymmetry  breaking with the ATLAS detector based on a disappearing-track signature in  pp collisions at $\sqrt{s}$ = 7 TeV}},  {\em Eur. Phys. J.} {\bf C72} (2012)  1993, [\href{http://xxx.lanl.gov/abs/1202.4847}{{\tt arXiv:1202.4847}}].

\bibitem{Aad:2008zzm}
{\bf ATLAS} Collaboration, {\it {The ATLAS Experiment at the CERN Large Hadron  Collider}},  {\em JINST} {\bf 3} (2008) S08003.

\bibitem{Paige:2003mg}
F.~E. Paige, S.~D. Protopopescu, H.~Baer, and X.~Tata, {\it {ISAJET 7.69: A  Monte Carlo event generator for $pp$, $\bar{p}p$, and $e^{+}e^{-}$  reactions}},  [\href{http://xxx.lanl.gov/abs/hep-ph/0312045}{{\tt  hep-ph/0312045}}].

\bibitem{Bahr:2008pv}
M.~Bahr, S.~Gieseke, M.~Gigg, D.~Grellscheid, K.~Hamilton, et~al., {\it  {Herwig++ Physics and Manual}},  {\em Eur. Phys. J.} {\bf C58} (2008)  639--707, [\href{http://xxx.lanl.gov/abs/0803.0883}{{\tt arXiv:0803.0883}}].

\bibitem{Sherstnev:2007nd}
A.~Sherstnev and R.~S. Thorne, {\it {Parton Distributions for LO Generators}},  {\em Eur. Phys. J.} {\bf C55} (2008) 553--575,  [\href{http://xxx.lanl.gov/abs/0711.2473}{{\tt arXiv:0711.2473}}].

\bibitem{Agostinelli:2002hh}
{\bf GEANT4} Collaboration, S.~Agostinelli et~al., {\it {GEANT4: A simulation  toolkit}},  {\em Nucl. Instrum. Meth.} {\bf A506} (2003) 250--303.

\bibitem{Aad:2010wqa}
{\bf ATLAS} Collaboration, {\it {The ATLAS Simulation Infrastructure}},  {\em  Eur. Phys. J.} {\bf C70} (2010) 823--874,  [\href{http://xxx.lanl.gov/abs/1005.4568}{{\tt arXiv:1005.4568}}].

\bibitem{Beenakker:1996ch}
W.~Beenakker, R.~Hopker, M.~Spira, and P.~Zerwas, {\it {Squark and gluino  production at hadron colliders}},  {\em Nucl. Phys.} {\bf B492} (1997)  51--103, [\href{http://xxx.lanl.gov/abs/hep-ph/9610490}{{\tt  hep-ph/9610490}}].

\bibitem{Kramer:2012bx}
M.~Kramer, A.~Kulesza, R.~van~der Leeuw, M.~Mangano, S.~Padhi, et~al., {\it  {Supersymmetry production cross sections in pp collisions at $\sqrt{s} =  7$~TeV}},  [\href{http://xxx.lanl.gov/abs/1206.2892}{{\tt arXiv:1206.2892}}].

\bibitem{ATLAS-CONF-2010-069}
{{\bf ATLAS} Collaboration}, {\it {Performance of primary vertex reconstruction  in proton-proton collisions at $\sqrt{s}=$7 TeV in the ATLAS experiment}},  ATLAS-CONF-2010-06, \url{http://cdsweb.cern.ch/record/128134}.

\bibitem{Cacciari:2008gp}
M.~Cacciari, G.~P. Salam, and G.~Soyez, {\it The anti-$k_t$ jet clustering  algorithm},  {\em JHEP} {\bf 04} (2008) 063,  [\href{http://xxx.lanl.gov/abs/0802.1189}{{\tt arXiv:0802.1189}}].

\bibitem{Aad:2011he}
{\bf ATLAS} Collaboration, {\it {Jet energy measurement with the ATLAS detector  in proton-proton collisions at $\sqrt{s}$ = 7 TeV}},  [\href{http://xxx.lanl.gov/abs/1112.6426}{{\tt arXiv:1112.6426}}], submitted to  {\em Eur. Phys. J.}

\bibitem{Aad:2011mk}
{\bf ATLAS} Collaboration, {\it {Electron performance measurements with the  ATLAS detector using the 2010 LHC proton-proton collision data}},  {\em Eur.  Phys. J.} {\bf C72} (2012) 1909,  [\href{http://xxx.lanl.gov/abs/1110.3174}{{\tt arXiv:1110.3174}}].

\bibitem{Aad:2011fp}
{\bf ATLAS} Collaboration, {\it {Measurement of the Transverse Momentum  Distribution of $W$ Bosons in pp Collisions at $\sqrt{s}$ = 7 TeV with the  ATLAS Detector}},  {\em Phys. Rev.} {\bf D85} (2012) 012005,  [\href{http://xxx.lanl.gov/abs/1108.6308}{{\tt arXiv:1108.6308}}].

\bibitem{Aad:2012re}
{\bf ATLAS} Collaboration, {\it {Performance of Missing Transverse Momentum  Reconstruction in Proton-Proton Collisions at 7 TeV with ATLAS}},  {\em Eur.  Phys. J.} {\bf C72} (2012) 1844,  [\href{http://xxx.lanl.gov/abs/1108.5602}{{\tt arXiv:1108.5602}}].

\bibitem{ATLAS:0lep7tev5ifb}
{\bf ATLAS} Collaboration, {\it {Search for squarks and gluinos with the ATLAS  detector in final states with jets and missing transverse momentum using 4.7  fb$^{-1}$ of $\sqrt{s} = 7$ TeV proton-proton collision data}},  [\href{http://xxx.lanl.gov/abs/1208.0949}{{\tt arXiv:1208.0949}}], submitted to  {\em Phys. Rev.} {\bf D}.

\bibitem{Nakamura:2010zzi}
{\bf Particle Data Group}, K.~Nakamura et~al., {\it {Review of  particle physics}},  {\em J. Phys.} {\bf G37} (2010) 075021. Corresponding  data may be found at pages 398--399.

\bibitem{Alwall:2011uj}
J.~Alwall, M.~Herquet, F.~Maltoni, O.~Mattelaer, and T.~Stelzer, {\it {MadGraph  5 : Going Beyond}},  {\em JHEP} {\bf 1106} (2011) 128,  [\href{http://xxx.lanl.gov/abs/1106.0522}{{\tt arXiv:1106.0522}}].

\bibitem{Sjostrand:2006za}
T.~{Sj\"{o}strand}, S.~Mrenna, and P.~Z. Skands, {\it {PYTHIA 6.4 Physics and  Manual}},  {\em JHEP} {\bf 0605} (2006) 026,  [\href{http://xxx.lanl.gov/abs/hep-ph/0603175}{{\tt hep-ph/0603175}}].

\bibitem{Alwall:2007fs}
J.~Alwall, S.~Hoche, F.~Krauss, N.~Lavesson, L.~Lonnblad, et~al., {\it  {Comparative study of various algorithms for the merging of parton showers  and matrix elements in hadronic collisions}},  {\em Eur. Phys. J.} {\bf C53}  (2008) 473--500, [\href{http://xxx.lanl.gov/abs/0706.2569}{{\tt  arXiv:0706.2569}}].

\bibitem{Aad:2011qe}
{\bf ATLAS} Collaboration, {\it {Measurements of underlying-event properties  using neutral and charged particles in $pp$ collisions at 900 GeV and 7 TeV  with the ATLAS detector at the LHC}},  {\em Eur. Phys. J.} {\bf C71} (2011)  1636, [\href{http://xxx.lanl.gov/abs/1103.1816}{{\tt arXiv:1103.1816}}].

\bibitem{Aad:2011dr}
{\bf ATLAS} Collaboration, {\it {Luminosity Determination in pp Collisions at  $\sqrt{s}$ = 7 TeV Using the ATLAS Detector at the LHC}},  {\em Eur. Phys.  J.} {\bf C71} (2011) 1630, [\href{http://xxx.lanl.gov/abs/1101.2185}{{\tt  arXiv:1101.2185}}].

\bibitem{LuminosityGroup:1367408}
{{\bf ATLAS} Collaboration}, {\it {Luminosity Determination in pp Collisions at  $\sqrt{s}$ = 7 TeV using the ATLAS Detector in 2011}},  ATLAS-CONF-2011-116,  \url{http://cdsweb.cern.ch/record/1376384}.

\bibitem{Cowan:2010js}
G.~Cowan, K.~Cranmer, E.~Gross, and O.~Vitells, {\it {Asymptotic formulae for  likelihood-based tests of new physics}},  {\em Eur. Phys. J.} {\bf C71}  (2011) 1554, [\href{http://xxx.lanl.gov/abs/1007.1727}{{\tt  arXiv:1007.1727}}].

\bibitem{Read:2002hq}
A.~L. Read, {\it {Presentation of search results: The CL(s) technique}},  {\em  J. Phys.} {\bf G28} (2002) 2693--2704.

\bibitem{Chen:1999yf}
C.~Chen, M.~Drees, and J.~Gunion, {\it {Addendum/erratum for `searching for  invisible and almost invisible particles at $e^{+}e^{-}$ colliders'  [hep-ph/9512230] and `a nonstandard string/SUSY scenario and its  phenomenological implications' [hep-ph/9607421]}},  [\href{http://xxx.lanl.gov/abs/hep-ph/9902309}{{\tt hep-ph/9902309}}].

\bibitem{Heister:2002mn}
{\bf ALEPH} Collaboration, A.~Heister et~al., {\it {Search for charginos nearly  mass degenerate with the lightest neutralino in $e^{+}e^{-}$ collisions at  center-of-mass energies up to 209 GeV}},  {\em Phys. Lett.} {\bf B533} (2002)  223--236, [\href{http://xxx.lanl.gov/abs/hep-ex/0203020}{{\tt  hep-ex/0203020}}].

\bibitem{Abbiendi:2002vz}
{\bf OPAL} Collaboration, G.~Abbiendi et~al., {\it {Search for nearly mass  degenerate charginos and neutralinos at LEP}},  {\em Eur. Phys. J.} {\bf C29}  (2003) 479--489, [\href{http://xxx.lanl.gov/abs/hep-ex/0210043}{{\tt  hep-ex/0210043}}].

\bibitem{Abdallah:2003gv}
{\bf DELPHI} Collaboration, J.~Abdallah et~al., {\it {Search for SUSY in the  AMSB scenario with the DELPHI detector}},  {\em Eur. Phys. J.} {\bf C34}  (2004) 145--156, [\href{http://xxx.lanl.gov/abs/hep-ex/0403047}{{\tt  hep-ex/0403047}}].

\bibitem{LEP2}
{LEP2 SUSY Working Group},  \url{http://lepsusy.web.cern.ch/lepsusy/www/inoslowdmsummer02/charginolowdm_pub.html}.

\bibitem{Adriani:2008zr}
{\bf PAMELA} Collaboration, O.~Adriani et~al., {\it {An anomalous positron  abundance in cosmic rays with energies 1.5-100 GeV}},  {\em Nature} {\bf 458}  (2009) 607--609, [\href{http://xxx.lanl.gov/abs/0810.4995}{{\tt  arXiv:0810.4995}}].

\bibitem{FermiLAT:2011ab}
{\bf Fermi LAT} Collaboration, M.~Ackermann et~al., {\it {Measurement of  separate cosmic-ray electron and positron spectra with the Fermi Large Area  Telescope}},  {\em Phys. Rev. Lett.} {\bf 108} (2012) 011103,  [\href{http://xxx.lanl.gov/abs/1109.0521}{{\tt arXiv:1109.0521}}].

\bibitem{Jarosik:2010iu}
N.~Jarosik et~al., {\it {Seven-Year Wilkinson Microwave Anisotropy Probe (WMAP)  Observations: Sky Maps, Systematic Errors, and Basic Results}},  {\em  Astrophys. J. Suppl.} {\bf 192} (2011) 14,  [\href{http://xxx.lanl.gov/abs/1001.4744}{{\tt arXiv:1001.4744}}].

\end{thebibliography}
\end{document}